\documentclass[12pt]{article}
\usepackage[dvips]{graphicx}
\begin{document}
\title{The damped Pinney equation and its applications to dissipative quantum mechanics}
\author{F Haas \footnote{Also at Universidade do Vale do Rio dos Sinos - UNISINOS, Brazil.}}
\date{\relax}
\maketitle
\begin{center}
{Institut f\"ur Theoretische Physik IV, 
Ruhr-Universit\"at Bochum\\ D-44780 Bochum, Germany}
\end{center}

\begin{abstract}
\noindent
The work considers the damped Pinney equation, defined as the model arising when a linear in velocity damping term is included in the Pinney equation. In the general case the resulting equation does not admit Lie point symmetries or is reducible to a simpler form by any obvious coordinate transformation. In this context the method of Kuzmak-Luke is applied to derive a perturbation solution, for weak damping and slow time-dependence of the frequency function. The perturbative and numerical solutions are shown to be in good agreement. The results are applied to examine the time-evolution of Gaussian shaped wave-functions in the Kostin formulation of dissipative quantum mechanics.
\end{abstract}


\maketitle

\section{Introduction}
As is well-known, the Pinney equation \cite{pinney} is ubiquitous in nonlinear dynamics. A partial list of applications includes the exact solution for the classical and quantum harmonic oscillators \cite{lewis1, lewis2}, the search for invariants (constants of motion) \cite{ermakov, ray, reid}, the stability analysis of charged particle motion in accelerators \cite{courant, qin}, the propagation of gravitational waves \cite{geralico}, the amplitude-phase representation of quantum mechanics \cite{milne}, the derivation of the Feynman propagator for variable-mass problems \cite{nassar}, numerical solutions for non relativistic quantum problems \cite{korsch}, cosmological particle-crea\-tion models \cite{ray2}, cosmological models for the Friedmann-Robertson-Walker metric \cite{williams}, isotropic, four-dimensional cosmological theories \cite{lidsey, hawkins}, ro\-ta\-ting shallow water-wave systems \cite{rogers}, curve flows in affine geometries \cite{qu}, the stabilizer set of Virasoro orbits \cite{guha}, Bose-Einstein condensates with time-dependent traps and/or time-dependent scattering length \cite{haas, herring}, discretized Pinney models \cite{common, schief} and nonlinear oscillations of transversally isotropic hyperelastic tubes \cite{mason}. 

However, in spite of such a large number of applications, the Pinney equation in itself does not have any dissipation term. Hence it is natural to generalize the model by inclusion of the simplest damping mechanism, a term li\-near in the velocity. We call the resulting model the {\it damped Pinney equation}. As discussed in more detail in Section V, damped Pinney equations arises in quantum mechanical models with dissipation. Specifically the Kostin formulation \cite{kostin} of the non conservative quantum time-dependent harmonic oscillator admits an exact Gaussian solution in terms of the solutions of a damped Pinney equation \cite{nassar3}. A damped Pinney equation also comes for Kostin's dissipative quantum mechanics under an arbitrary potential, after an expansion in the neighborhood of the classical path \cite{nassar2}. In addition, the Pauli equation for the Aharonov-Bohm effect with a time-dependent mass particle also reduces to a damped Pinney equation \cite{bouguerra}. 

There is a close relation between the solutions of the time-dependent harmonic oscillator and of the Pinney equation, see the nonlinear superposition law (\ref{s}) at Section II. At first sight one can expect some simple connection between the solutions of the damped harmonic oscillator and damped Pinney equations. However, as is shown in the following, this relationship, if existing, is non trivial. Additional difficulties arises from the non existence of useful obvious point or nonlocal transformations removing the damping term and casting the damped Pinney equation into some well-known integrable class of equations. Reduction to Abel or Emden-Fowler equations can be shown to be trivial, but this does not help very much in solving the original problem. Indeed such classes of equations can be integrated only in some particular cases but not for the general damped Pinney equation. Finally the symmetry structure of the damped Pinney equation is poor in comparison to the three-parameter group, $SL(2,R)$, always admitted by the usual Pinney equation. These issues are discussed in more detail in Section II. 

In view of the above difficulties the present work is dedicated to the mo\-dest task of providing an approximate solution for the damped Pinney equation, assuming weak damping. In addition a slow variation of the frequency function is allowed. The perturbation technique chosen is the method of Kuzmak-Luke \cite{kevorkian, kuzmak, luke}, which is appropriate to generate bounded solutions for strongly nonlinear oscillator equations.  Afterwards the perturbative solution is applied to dissipative quantum mechanics in Kostin's version. A damped Pinney equation appears naturally in this formulation of dissipative quantum systems. 

This work is organized as follows. In Section II the damped Pinney equation is presented and some of its basic properties are discussed. In Section III the Kuzmak-Luke method is applied to derive a perturbative solution, free from secular divergences. Section IV shows good agreement of the approximate solution with some numerical examples. In Section V the general connexion between the damped Pinney equation and dissipative quantum mechanics is explored. The final Section is dedicated to the conclusions.  

\section{The damped Pinney model}
As is well known \cite{pinney}, the general solution for the Pinney equation 
\begin{equation}
\label{e0}
\ddot{x} + \omega^{2}(t) x = \frac{k}{x^3} \,, 
\end{equation}
where $\omega = \omega(t)$ is a time-dependent frequency function and $k$ a numerical constant, can be written as 
\begin{equation}
\label{s}
x = (c_1 \sigma_{1}^2 + c_2 \sigma_{2}^2 + 2 c_3 \sigma_{1}\sigma_{2})^{1/2} \,,
\end{equation}
where $c_{1}, c_{2}$ and $c_{3}$ are constants such that $c_{1}c_{2} - c_{3}^2 = k$ and $\sigma_1$ and $\sigma_2$ are solutions for the time-dependent harmonic oscillator equation, 
\begin{equation}
\ddot{\sigma}_i + \omega^{2}(t) \sigma_i = 0 \,, \quad i=1, 2\,,
\end{equation}
with unit Wronskian, $\sigma_{1}\dot\sigma_2 - \sigma_{2}\dot\sigma_1 = 1$. We restrict $k$ to be positive to prevent ``collapse into the origin" issues. Notice the differences to the quantum case, where the wave-function for the time-dependent singular harmonic oscillator is regular provided $k > - \hbar^2/(4m^2)$, restoring dimensional quantities \cite{camiz}. However, in the formal classical limit, setting $\hbar \equiv 0$ we strictly need $k > 0$ to avoid $x$ collapsing to the origin. This can be verified {\it e.g.} from the solution shown in Eq. (\ref{s}) or from the potential function associated to Eq. (\ref{e0}), namely $V = \omega^2 x^2/2 + k/(2 x^2)$ which is not bounded from below if $k < 0$. As a particular example for $\omega = 1$ and initial conditions $x(0) = 1, \dot{x}(0) = 0$, Eq. (\ref{s}) gives $x^2 = \cos^2 t + k\sin^2 t$. Therefore one would have $x = 0$ at some time $t = t_{*} > 0$ for any $k \leq 0$.

Equation (\ref{e0}) does not include any mechanism for dissipation. In this context it is natural to add a term linear in the velocity, yielding the damped Pinney equation 
\begin{equation}
\label{e1}
\ddot{x} + 2\epsilon\dot{x} + \omega^{2}(t) x = \frac{k}{x^3} \,, 
\end{equation}
where  $\epsilon > 0$ is a constant positive parameter. Notice that a time-dependence of the damping coefficient could be easily removed by an appropriate change of coordinates. Hence the basic properties of the system are already displayed for a constant $\epsilon$. 

Unlike the undamped ($\epsilon \equiv 0$) case the integration of Eq. (\ref{e1}) is a challenge in general. Indeed Eq. (\ref{e1}) does not possess an universal Lie point symmetry group for arbitrary frequency function. This is in contrast to the richer $sl(2,R)$ algebra always admitted by the usual Pinney equation \cite{nucci} for any choice of $\omega(t)$. It can be shown that the damped Pinney equation has geometric symmetries for  particular functional dependencies of $\omega(t)$, but these special cases are outside the scope of the present work. Also the application of symmetry generators which are quadratic in the velocity does not produce new results. Actually a coordinate transformation put Eq. (\ref{e1}) into the form of a generalized Emden-Fowler equation of index $-3$. There is an extensive literature on the integrability of generalized Emden-Fowler equations of arbitrary index \cite{conte, govinder}, but we believe that it is still interesting to investigate in more detail the index $n = -3$ case. 

There is no easily identifiable point transformation, so that the damped Pinney equation could be put into a simpler, always integrable form. As a tentative example, consider the quasi-invariance transform 
\begin{equation}
x = \rho(t) Q(T) \,, \quad T = T(t) \,,
\end{equation}
where $Q$ is the new dependent variable and $\rho, T$ are functions of time to be determined according to convenience. The resulting equation is given by 
\begin{equation}
\label{fu}
\rho \dot{T}^2 \,\frac{d^2 Q}{dT^2} + (\rho \ddot{T} + 2\dot\rho \dot{T} + 2\epsilon\rho\dot{T}) \,\frac{dQ}{dT} + (\ddot\rho + \omega^2 \rho)\,Q = \frac{k}{\rho^3 Q^3} \,.
\end{equation}
From Eq. (\ref{fu}), it is easy to verify that it is not possible simultaneously to eliminate the damping term and  set the coefficient of the inverse cubic term to a constant. 

Indeed the damping term in Eq. (\ref{fu}) can be eliminated provided 
\begin{equation}
\label{fuu}
\dot{T} = \frac{e^{-2\epsilon t}}{\rho^2} \,,
\end{equation}
ignoring an irrelevant multiplicative constant which could be included. By use of Eqs. (\ref{fu}--\ref{fuu}) we derive   
\begin{equation}
\label{e3}
\frac{d^2 Q}{dT^2} + W^2 Q = \frac{k e^{4\epsilon t}}{Q^3} \,,
\end{equation}
where $W = W(t)$ is a function of time, defined in terms of the auxiliary equation
\begin{equation}
\label{e4}
\ddot\rho + 2\epsilon\dot\rho + \omega^2 \rho = \frac{W^2 e^{-4\epsilon t}}{\rho^3} \,.
\end{equation}

One can choose $W = {\rm constant}$ or even $W = 0$, but in all cases Eq. (\ref{e3}) would be still explicitly time-dependent through the inverse cubic term, since $\epsilon \neq 0$ and $k$ is assumed constant. The non autonomous character of Eq. (\ref{e3}) prevents solvability in general. The time-dependence should be written using $t = t(T)$, obtained via Eq. (\ref{fuu}).

At least, setting $W \equiv 0$, one can cast the problem in the form of a generalized Emden-Fowler equation of index $-3$,
\begin{equation}
\label{ef}
\frac{d^2 Q}{dT^2}  = \frac{\mu(T)}{Q^3} \,,
\end{equation}
where $\mu(T) = k\exp(4\epsilon t)$. This is equivalent to a choice of gauge \cite{conte}. However, from the practical point of view of computing the solution, this property is not very helpful. Indeed consider the simplest case of a constant frequency function $\omega = \omega_0$. Since $\dot\omega = 0$, it is trivial to solve Eq. (\ref{e4}) assuming $W = 0$. Using Eq. (\ref{fuu}) and the particular solution $\rho = \exp(-\epsilon t) \cos[\sqrt{\omega_{0}^2 - \epsilon^{2}}\,t]$, Eq. (\ref{e3}) becomes
\begin{equation}
\label{p}
\frac{d^2 Q}{dT^2} = \frac{k}{Q^3}\exp\Bigg(\frac{4\epsilon}{\sqrt{\omega_{0}^2 - \gamma^2}} \tan^{-1}\Bigl[\sqrt{\omega_{0}^2 - \gamma^2}\,\, T\Bigr]\Bigg) \,.
\end{equation}
Equation (\ref{p}) does not fall into any of the known integrable Emden-Fowler equations \cite{govinder, berkovich}. Evidently, trying with different particular solutions for Eq. (\ref{e4}) does not improve the scenario. 

For time-dependent frequencies the same procedure applies, but the resulting Emden-Fowler equation could be even more complicated.  In the undamped ($\epsilon \equiv 0$) case, Eq. (\ref{p}) reduces to the Ermakov-Pinney \cite{pinney, ermakov} equation, the integrability of which is well-known. Also in general the non autonomous Emden-Fowler equation of index $-3$ does not possess the Painlev\'e property. For instance Refs. \cite{conte, govinder} consider only the situation in which $\mu$ is a constant. In addition there is no quadratic constant of motion for Eq. (\ref{e1}), except in the autonomous case \cite{conte}.

A detailed analysis shows that the damped Pinney equation does not matches the Tresse-Cartan conditions \cite{tresse, cartan} and hence is not linearizable through a general point transformation. Also generalized Sundman transformations $Q = Q(x,t), \,dT = F(x,t)\,dt, \,F\partial Q/\partial x \neq 0$, can be shown to be useless. Indeed the damped Pinney equation does not matches the conditions established in \cite{duarte, euler} in order to be reduced to the free particle or similar simple equations via generalized Sundman transformations. 

Another avenue could be the reduction to an Abel equation of the second kind through $v \equiv v(x) = \dot{x}$, so that 
\begin{equation}
\label{ab}
v\frac{dv}{dx} = - 2\epsilon v - \omega^2 x + \frac{k}{x^3} \,.
\end{equation}
However, Eq. (\ref{ab}) does not correspond to any of the known integrable Abel equations \cite{boyko, cheb, polyanin}. Also certain classes of nonlocal symmetries  are not helpful, see Eq. (5.10) of Ref. \cite{adam}. The difficulties are due to the singular $\sim x^{-3}$ term.

Some Pinney equations with ti\-me-de\-pen\-dent damping and nonlinear terms have been considered in the literature, e.g. in the calculation of the geo\-me\-tric phases and angles of dynamical systems. Specifically \cite{maamache} these modified Pinney equations are given by  
\begin{equation}
\label{m}
\ddot{x} - \frac{\dot{m}}{m}\,\dot{x} + \omega^2 x = \frac{k m^2}{x^3} \,,
\end{equation}
where $m$, which can be interpreted as a time-dependent mass, and $\omega$ are arbitrary functions of time and $k$ is a constant. Notice that in Eq. (\ref{m}) the form of the time-dependence allows the use of 
\begin{equation}
\label{t}
Q = \frac{x}{\sqrt{m}\,\,\rho} \,, \quad T = \int\frac{dt}{\rho^2} \,, 
\end{equation}
where $\rho$ is an arbitrary function of time, to convert Eq. (\ref{m}) to a Pinney equation in standard form, 
\begin{equation}
\frac{d^2 Q}{dT^2} + \rho^3 \Bigl[\ddot{\rho} + (\omega^2 + \frac{\ddot m}{2m} - \frac{3\dot{m}^2}{4m^2})\rho\Bigr] Q = \frac{k}{Q^3} \,.
\end{equation}
Therefore these systems are somewhat trivial. Notice the particular form of the nonlinear term, which needs to be proportional to $m^2$ so that the scaling (\ref{t}) works. From a physical point of view, the most relevant case would be the one in which the nonlinear term is time-independent. In addition other, different classes of generalized Pinney models with nonlinear and time-dependent damping have been discussed elsewhere \cite{qian}. The conclusion of the Section is that apparently damped Pinney equations in the form (\ref{e1}) does not possess a closed form solution valid for general $\omega(t)$, unlike the standard Pinney equation. The same applies to the alternative form given by Eq. (\ref{ef}) for arbitrary $\mu(t)$. 

\section{Approximate solution for weak damping and slow time-dependence}
The difficulties enumerated in the last Section suggest the use of perturbation theory, in the case of weak damping. In other words we restrict to a small damping coefficient, $\epsilon$. An approximate solution would be welcome for physical applications as shown in Section IV. In this manuscript  this modest goal of deriving an approximate solution is achieved through the method of Kuzmak-Luke \cite{kevorkian, kuzmak, luke}, which is good enough for equations which remain nonlinear even when the perturbation parameter goes to zero. Indeed this is the case of Eq. (\ref{e1}) when $\epsilon \rightarrow 0$. In addition the Kuzmak-Luke technique applies only for slowly varying non-autonomous systems. Therefore we consider general frequency functions of the form 
\begin{equation}
\label{slow}
\omega = \Omega(\epsilon t) \,,
\end{equation}
so that the time-dependence is slow. In Eq. (\ref{slow}), $\Omega$ is an arbitrary analytic function of its argument. To avoid some singularities it is assumed that $\Omega > 0$.  Fortunately Eq. (\ref{e1}) in the non perturbed ($\epsilon \equiv 0$) case has a periodic solution so that all conditions to apply the Kuzmak-Luke method are satisfied. Naive perturbation theory would lead to a divergent solution. 

Other methods of removal of the divergences in the perturbation series, like the Lindstedt-Poincar\'e approach, could also be chosen. However, we verified that the insistence on the use of such variational methods just adds unnecessary technical problems, even observing that Eq. (\ref{e1}) admits the Lagrange function
\begin{equation}
\label{e5}
L = L(x,\dot{x},t) = \frac{e^{2\gamma t}}{2}(\dot{x}^2 - \omega^2 x^2 - \frac{k}{x^2}) \,.
\end{equation}
In practice it can be checked that the Kuzmak-Luke method is among the simplest reliable approaches for the problem. 

The Kuzmak-Luke procedure seeks a series solution 
\begin{equation}
x = x_{0}(\tau,\tilde{t}) + \epsilon x_{1}(\tau,\tilde{t}) + \dots \,,
\end{equation}
where $\tilde{t} = \epsilon t$ and
\begin{equation}
\frac{d\tau}{dt} = f(\tilde{t}) \,,
\end{equation}
with $f = f(\tilde{t})$ determined by the requirement that $x_0$ be periodic in $\tau$ with a constant period which can be taken as $2\pi$ without loss of generality. In the calculations $\tau$ and $\tilde{t}$ are regarded as independent variables. 

To zeroth-order in $\epsilon$, 
\begin{equation}
\label{z}
f^{2}(\tilde{t}) \frac{\partial^2 x_0}{\partial\tau^2} + \Omega^{2}(\tilde{t}) x_0 = \frac{k}{x_{0}^3} \,,
\end{equation}
with no damping term. Equation (\ref{z}) can be integrated once so that we obtain the slowly varying energy function 
\begin{equation}
\label{en}
E_0 = E_{0}(\tilde{t}) = \frac{f^2}{2}\left(\frac{\partial x_0}{\partial\tau}\right)^2 + \frac{\Omega^2 x_{0}^2}{2} + \frac{k}{2x_{0}^2} \,.
\end{equation}
The energy integral can be used to derive the quadrature
\begin{equation}
\label{x0}
x_0 = \pm \left[\left(\frac{k}{\Omega^{2}(\tilde{t})} + A^{4}(\tilde{t})\right)^{1/2} + A^{2}(\tilde{t})\,\cos\left(\frac{2\,\Omega(\tilde{t})\tau}{f(\tilde{t})} + \phi(\tilde{t})\right)\right]^{1/2} \,,
\end{equation}
where the slowly varying amplitude $A = A(\tilde{t})$ and phase $\phi = \phi(\tilde{t})$ are chosen according to convenience. For simplicity we ignore the time-dependence of the phase, setting $\phi = {\rm constant}$ In terms of the amplitude the energy is $E_0 = \Omega \sqrt{\Omega^2 A^4 + k}$, reproducing the usual law $E_0 \sim A^2$ in the linear ($k \equiv 0$) case. From now we adopt the positive sign in Eq. (\ref{x0}).

By inspection, the trajectories defined by Eq. (\ref{x0}) become periodic with period $2\pi$, independent on $\tilde{t}$, if we define
\begin{equation}
\label{f}
f(\tilde{t}) = 2\,\Omega(\tilde{t}) \,,
\end{equation}

We should determine $A^{2}(\tilde{t})$ requiring the first-order correction $x_1$ to be a periodic function of $\tau$, free from mixed secular terms. In the present case this can be shown to be equivalent to 
\begin{equation}
\label{i}
\frac{\exp(2\,\tilde{t})}{\Omega(\tilde{t})}\,\int_{0}^{2\pi} d\tau \left(E_{0}(\tilde{t}) - \frac{\Omega^{2}(\tilde{t})\, x_{0}^{2}(\tau,\tilde{t})}{2} - \frac{k}{2x_{0}^{2}(\tau,\tilde{t})}\right) = {\rm constant} \,,
\end{equation}
where the integration is performed for fixed $\tilde{t}$. For more details see for instance Eq. (3.6.35a) of Ref. \cite{kevorkian}. Eq. (\ref{i}) gives 
\begin{equation}
\label{a2}
A^2 = \frac{\sqrt{\Omega_0}\, A_0 \exp(-\tilde{t})}{\Omega(\tilde{t})} \, \left(\sqrt{k} + \frac{\Omega_{0}A_{0}^2 \exp(-2\tilde{t})}{4}\right)^{1/2} \,,
\end{equation}
where $\Omega_0 \equiv \Omega(0)$ and $A_0$ is a reference value. 

Finally the combination of Eqs. (\ref{x0}), (\ref{f}) and (\ref{a2}) gives the zeroth-order solution 
\begin{eqnarray}
\label{fi}
x_0 &=& \frac{1}{\sqrt{\Omega(\epsilon t)}}\,\Biggl[\sqrt{k} + \frac{\Omega_{0}A_{0}^{2} e^{-2\epsilon t}}{2} \\ &+& \sqrt{\Omega_{0}}A_{0}\, e^{-\epsilon t}\,\left(\sqrt{k} + \frac{\Omega_{0}A_{0}^{2} e^{-2\epsilon t}}{4}\right)^{1/2}\cos\left(2\int_{t_0}^{t}\Omega(\epsilon t')\,dt'\right)\Biggr]^{1/2} \,, \nonumber
\end{eqnarray}
containing the two constants of integration $A_0 > 0$ and $t_0$. The perturbation procedure could be carried out to higher orders, but Eq. (\ref{fi}) is sufficient for our purposes. 

The relevance of equation (\ref{fi}) cannot be overestimated. It provides a sort of JWKB solution for the damped Pinney equation, reproducing the qua\-li\-ta\-ti\-ve properties which are expected, namely, it shows a periodic motion spiraling toward the slowly varying ``fixed" point $k^{1/4}/\,\Omega^{1/2}(\epsilon t)$. In addition it can be checked that $x_0 > 0$ for all times provided $k > 0$. When there is no damping ($\epsilon \rightarrow 0$) nor  nonlinearity ($k \rightarrow 0$), we derive $x_0 \rightarrow A_0 \cos[\Omega_0 (t-t_{0})]$ as it should be. Of course one can adopt the simplistic viewpoint that a numerical solution could be sufficient so that the expression (\ref{fi}) is irre\-le\-vant. This na\"ive criticism does not appreciate the usefulness of exact or approximate analytical solutions in general. However, it is clear that expressions like (\ref{fi}) provide information not so easily found from numerics, as, for instance, the functional dependence of the solutions on the amplitude $A_0$. 

If there were no damping, but the frequency could still depends on $\tilde{t}$, Eq. (\ref{fi}) would be replaced by 
\begin{equation}
\label{fii}
x_0 = \frac{1}{\sqrt{\Omega(\epsilon t)}}\,\Biggl[\sqrt{k} + \frac{\Omega_{0}A_{0}^{2}}{2} + \sqrt{\Omega_{0}}A_{0}\!\left(\sqrt{k} + \frac{\Omega_{0}A_{0}^{2}}{4}\right)^{1/2}\!\!\!\!\!\!\cos\left(2\!\int_{t_0}^{t}\Omega(\epsilon t')\,dt'\right)\Biggr]^{1/2} \,. 
\end{equation}
This is the form arising from the well-known exact solution (\ref{s}) of the standard Pinney equation, considering the JWKB solutions $\sigma_1 = \Omega^{-1/2} \, \cos[\int\Omega \, dt']$ and $\sigma_2 = \Omega^{-1/2} \, \sin[\int\Omega \, dt']$ for the time-dependent harmonic oscillator along with appropriate parameters $c_i$. The powerfulness of both Eqs. (\ref{fi}) and (\ref{fii}) is in their generality: they provide approximate solutions for the damped or standard Pinney equations, irrespective of the functional form of the frequency, as long as it is slowly varying and positive. Near singular times, at which $\Omega = 0$, the solutions would clearly be not appropriate. 

It is interesting to observe that
\begin{equation}
E_0 = \Omega \left(\frac{\Omega_0 A_{0}^2 e^{-2\epsilon t}}{2} + \sqrt{k}\right) \,,
\end{equation}
as follows from Eqs. (\ref{en}) and (\ref{fi}), generalizing the usual adiabatic theorem to the damped and nonlinear case. It is also consistent with the asymptotic approaching toward the ``fixed" point $k^{1/4}/\,\Omega^{1/2}$, where $E_0 = \Omega\sqrt{k}$. 

\section{Simple examples}
We have compared the approximate solution (\ref{fi}) with numerical simulations for specific frequency functions as follows. 

\subsection{$\Omega = \Omega_0$}
For a constant frequency, $\Omega = \Omega_0$, it is expected that the trajectory approaches the value $k^{1/4}/\,\Omega_{0}^{1/2}$. Figure 1 shows the perturbative solution (\ref{fi}) for the parameters $\epsilon = 0.1, \Omega_0 = k = 1, A_0 = 2$ and $t_0 = 0$. For the parameters, one found $x_{0}(0) = 2.41, \dot{x}_{0}(0) = -0.17$. Simulation of the corresponding damped Pinney equation with these initial conditions fully confirms the accuracy of the approximate solution with no noticeable difference to Fig. 1. 

\begin{figure*}
\begin{center}
\includegraphics{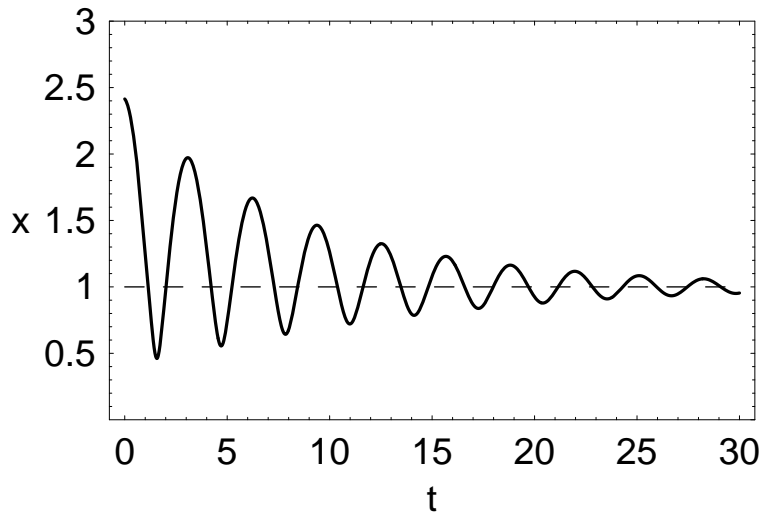}
\caption{Approximate solution (\ref{fi}) for $\Omega = \Omega_0$ and parameters $\epsilon = 0.1, \Omega_0 = k = 1, A_0 = 2$ and $t_0 = 0$. Accordingly, the initial conditions are $x_{0}(0) = 2.41, \dot{x}_{0}(0) = -0.17$. Direct numerical simulation reproduces the same curve.}
\label{figure1}
\end{center}
\end{figure*}  

\subsection{$\Omega = \Omega_{0}\,(1+\tilde{t}^{\,2}\,)^{-1/2}$}
For a decaying frequency, $\Omega = \Omega_{0}\,(1+\tilde{t}^{\,2}\,)^{-1/2}$, the trajectory shows the asymptotic behavior  $x_0 \rightarrow \Omega_{0}^{-1/2} k^{1/4} (1+\tilde{t}^{\,2}\,)^{1/4}$ because the confining potential becomes weaker. Figure 2 shows the perturbative solution (\ref{fi}) for the same parameters as in Fig. 1. There is no need to show the numerical simulation, because it gives the same result as Fig. 2. 

\begin{figure*}
\begin{center}
\includegraphics{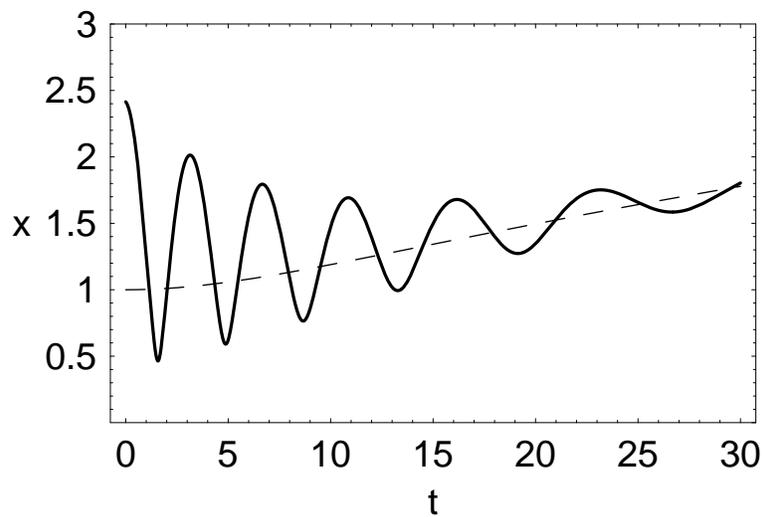}
\caption{Approximate solution (\ref{fi}) for $\Omega = \Omega_{0}\,(1+\epsilon^2 t^{2})^{-1/2}$ and parameters $\epsilon = 0.1, \Omega_0 = k = 1, A_0 = 2$ and $t_0 = 0$. Accordingly the initial conditions are $x_{0}(0) = 2.41, \dot{x}_{0}(0) = -0.17$. Direct numerical simulation reproduces the same curve. The dashed curve shows the asymptotic behavior $x \sim (1+\epsilon^2 t^{2})^{1/4}$.}
\label{figure2}
\end{center}
\end{figure*}  

\subsection{$\Omega = \Omega_{0}\,(1+\tilde{t}^{\,2}\,)^{1/2}$}
When the frequency grows as $\Omega = \Omega_{0}\,(1+\tilde{t}^{\,2}\,)^{1/2}$, the trajectory shows the asymptotic behavior  $x_0 \rightarrow \Omega_{0}^{-1/2} k^{1/4} (1+\tilde{t}^{\,2}\,)^{-1/4}$ because the confining potential becomes stronger. This is shown in Figure 3 with the same parameters as before. It reproduces very well the numerical simulation. 

\begin{figure*}
\begin{center}
\includegraphics{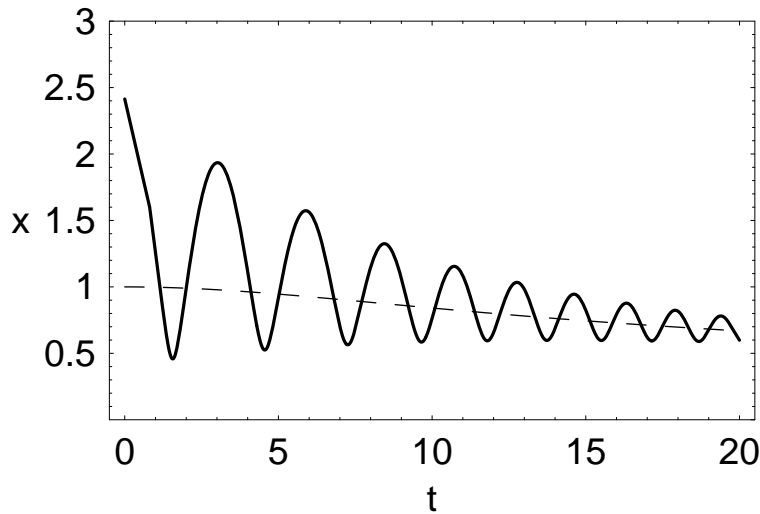}
\caption{Approximate solution (\ref{fi}) for $\Omega = \Omega_{0}\,(1+\epsilon^2 t^{2})^{1/2}$ and the same parameters of Figs. 1 and 2. The graphic is in accordance with the numerical results. The dashed curve shows the asymptotic behavior $x \sim (1+\epsilon^2 t^{2})^{-1/4}$.}
\label{figure3}
\end{center}
\end{figure*}  

To conclude the Section we have verified that the perturbation and numerical solutions disagree in the case of strong damping as can be expected. 

\section{Application to dissipative quantum mechanics}
A popular approach for the time-dependent dissipative quantum harmonic oscillator is given by Kostin's model \cite{kostin}, which is expressed in terms of the Kostin equation
\begin{equation}
\label{ks}
- \frac{\hbar^2}{2m}\frac{\partial^2 \psi}{\partial q^2} + \Bigg(\frac{m\omega^{2}(t)q^2}{2} + \frac{\hbar\epsilon}{i}\ln\left(\frac{\psi}{\psi^{*}}\right)\Bigg)\psi = i\hbar\frac{\partial\psi}{\partial t} \,.
\end{equation}
In Eq. (\ref{ks}) $\epsilon$ is the damping coefficient and the remaining symbols have their usual meaning. In particular the wave-function is $\psi = \psi(q,t)$. The Kostin nonlinear modification of the Schr\"odinger equation is among the only reliable alternatives to include damping in quantum mechanics, due to its smooth classical limit \cite{spiller}. In particular it is better than considering 
non-Hermitian terms in the Hamiltonian since a continuity equation follows from 
Eq. (\ref{ks}). 

Consider the de Broglie-Bohm decomposition $\psi = \sqrt{n(q,t)}\exp(iS(q,t)/\hbar)$, assuming the Gaussian {\it Ansatz} 
\begin{equation}
\label{nn}
n = \Bigl(\pi x^{2}(t)\Bigr)^{- 1/2} \exp\Bigl(-\Bigl(\frac{q - q_{cl}(t)}{x(t)}\Bigr)^{2}\Bigr) \,,
\end{equation}
where $x = x(t)$ and $q = q_{cl}(t)$ are functions to be determined and the quantum-mechanical fluid velocity is 
\begin{equation}
\label{k1}
u(q,t) = \frac{1}{m}\frac{\partial S}{\partial q} = \frac{\dot{x}}{x} (q - q_{cl}) + \dot{q}_{cl} \,. 
\end{equation}
From the Kostin equation it follows \cite{nassar3} that
\begin{eqnarray}
\label{k2}
\ddot{x} + 2\epsilon\dot{x} + \omega^{2}(t)x &=& \frac{\hbar^2}{m^2 x^3} \,,\\
\label{k3}
\ddot{q}_{cl} + 2\epsilon\dot{q}_{cl} + \omega^{2}(t)q_{cl} &=& 0 \,.
\end{eqnarray}
Clearly $x$ satisfy a damped Pinney equation and $q_{cl}$ solves the classical Newton equation. The same equations can be derived for arbitrary nonlinear potentials, ex\-pan\-ding around the classical trajectory \cite{nassar2}. Assuming weak damping and slowly varying frequencies, we can apply the results of  Section III. 

As an example consider the oscillating frequency
\begin{equation}
\omega = \Omega_0 (1 + \gamma \sin(2\epsilon t)) \,,
\end{equation}
where $\Omega_0 > 0$ and $0 < \gamma < 1$ are fixed parameters. Using the approximate solution (\ref{fi}) and numerically solving Eq. (\ref{k3}) with the initial condition $q_{cl}(0) = 1, \dot{q}_{cl}(0) = 0$, we derive both the quantum fluid density and velocity fields from Eqs. (\ref{nn}) and (\ref{k1}). The results are shown in Figs. 4 (the standard deviation of the wave-packet), 5 (the particle density) and 6 (the velocity field at fixed position $q = 0$). In these graphs the parameters are $\gamma = 0.7, \hbar = m = \Omega_0 = 1, \epsilon = 0.1, A_0 = 4, t_0 = 0$. Other examples can be easily constructed as well. 

\begin{figure*}
\begin{center}
\includegraphics{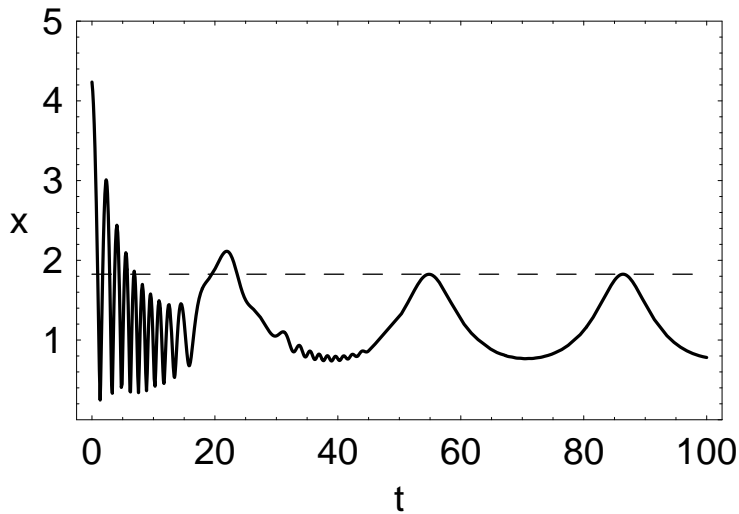}
\caption{Approximate solution for (\ref{k2}) with $\omega = \Omega_0 (1 + \gamma \sin(2\epsilon t))$ and the parameters $\gamma = 0.7, \hbar = m = \Omega_0 = 1, \epsilon = 0.1, A_0 = 4, t_0 = 0$.}
\label{figure4}
\end{center}
\end{figure*}  

\begin{figure*}
\begin{center}
\includegraphics{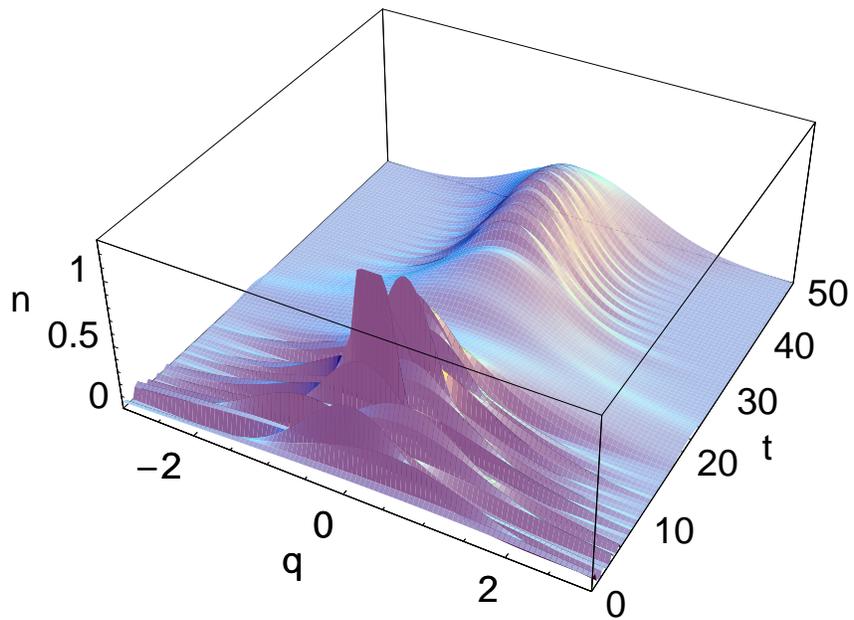}
\caption{The particle density as follows from Eq. (\ref{nn}) and the use of the approximate solution shown in Figure 4. Initial conditions for the auxiliary field: $q_{cl}(0) = 1, \dot{q}_{cl}(0) = 0.$ }
\label{figure5}
\end{center}
\end{figure*}  

\begin{figure*}
\begin{center}
\includegraphics{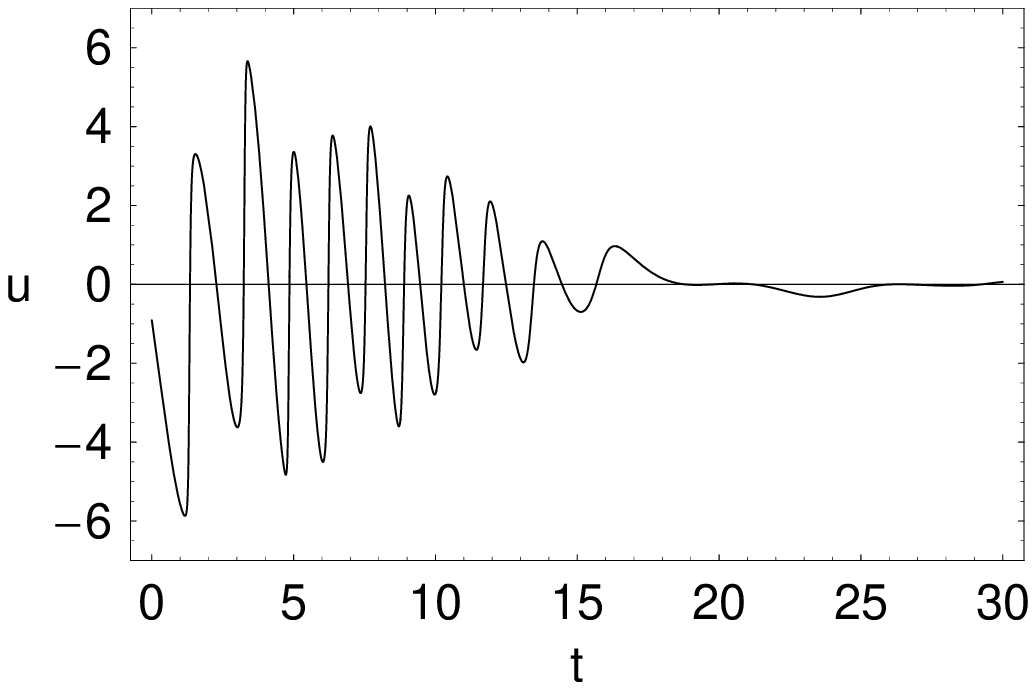}
\caption{The velocity field associated to the same parameters of Figs. 4 and 5, evaluated at $q = 0$.}
\label{figure6}
\end{center}
\end{figure*}

\section{Conclusion}
This manuscript derives an approximate solution for the autonomous damped Pinney equation for weak damping and slowly varying frequency using the Kuzmak-Luke method. The JWKB-like solution (\ref{fi}) reproduces with accuracy the numerical solutions of the model. Due to the usefulness of the Pinney equation in many areas of physics, mathematics and engineering, it would be relevant to derive general statements about the corresponding model when dissipation is present. In this context the perturbation approach here is just a first essay.  However, probably there is no closed-form solution for the damped Pinney equation, valid for arbitrary frequency functions, in contrast to the undamped case. Adopting the optimistic view that such an universal solution could be constructed, one can expect the need for more complex nonlocal mappings than the generalized Sundman transformations. As a by-product there would be further insight on fundamental questions about dissipative non autonomous quantum mechanics. Up to now only special classes of damped Pinney equations or the equivalent Emden-Fowler equation (\ref{ef}) are known to be amenable to reduction of order. The interesting point about the expression (\ref{fi}) is that the frequency function is not of a particular form, except for having a slow time-dependence and not becoming zero. More accurate results could be found extending the perturbation theory to higher orders. 

\vskip .5cm
\noindent
{\bf Acknowledgments:}\\{This work was sup\-por\-ted by the A\-le\-xan\-der von Humboldt Foundation. The author is grateful to Prof. J. M. F. Bassalo for suggesting the problem and for fruitful discussions.}

\end{document}